 \documentclass{ecai} 

\usepackage{latexsym}
\usepackage{amssymb}
\usepackage{amsmath}
\usepackage{amsthm}
\usepackage{booktabs}
\usepackage{enumitem}
\usepackage{graphicx}
\usepackage{color}
\usepackage{url}
\usepackage{hyperref}
\usepackage{subcaption}
\usepackage{caption}
\usepackage[breakable,skins]{tcolorbox}
\usepackage{multirow}
\usepackage{float}
\usepackage{lipsum}
\usepackage{wrapfig}
\usepackage{textcomp}
\usepackage{tabularx}

\newcommand{\BibTeX}{B\kern-.05em{\sc i\kern-.025em b}\kern-.08em\TeX}

\begin{document}

%%%%%%%%%%%%%%%%%%%%%%%%%%%%%%%%%%%%%%%%%%%%%%%%%%%%%%%%%%%%%%%%%%%%%%%%

\begin{frontmatter}

%%% Use this command to specify your submission number.
%%% In doubleblind mode, it will be printed on the first page.

\paperid{2477} 

%%% Use this command to specify the title of your paper.

\title{Meta-RAG on Large Codebases Using Code Summarization}

%%% Use this combinations of commands to specify all authors of your 
%%% paper. Use \fnms{} and \snm{} to indicate everyone's first names 
%%% and surname. This will help the publisher with indexing the 
%%% proceedings. Please use a reasonable approximation in case your 
%%% name does not neatly split into "first names" and "surname".
%%% Specifying your ORCID digital identifier is optional. 
%%% Use the \thanks{} command to indicate one or more corresponding 
%%% authors and their email address(es). If so desired, you can specify
%%% author contributions using the \footnote{} command.

\author[A]{\fnms{Vali}~\snm{Tawosi}\orcid{0000-0001-5052-672X}\thanks{Corresponding Author. Email: vail.tawosi@jpmorgan.com}}
\author[A]{\fnms{Salwa}~\snm{Alamir}\orcid{0009-0006-6650-7041}}
\author[B]{\fnms{Xiaomo}~\snm{Liu}\orcid{0000-0003-4184-4202}} 
\author[B]{\fnms{Manuela}~\snm{Veloso}\orcid{0000-0001-6738-238X}} 

\address[A]{JP Morgan AI Research, UK}
\address[B]{JP Morgan AI Research, US}

%%% Use this environment to include an abstract of your paper.

\begin{abstract}
    Large Language Model (LLM) systems have been at the forefront of applied Artificial Intelligence (AI) research in a multitude of domains.
    One such domain is software development, where researchers have pushed the automation of a number of code tasks through LLM agents. Software development is a complex ecosystem, that stretches far beyond code implementation and well into the realm of code maintenance.
    In this paper, we propose a multi-agent system to localise bugs in large pre-existing codebases using information retrieval and LLMs. Our system introduces a novel Retrieval Augmented Generation (RAG) approach, Meta-RAG, where we utilise summaries to condense codebases by an average of 79.8\%, into a compact, structured, natural language representation. We then use an LLM agent to determine which parts of the codebase are critical for bug resolution, i.e. bug localisation.
    We demonstrate the usefulness of Meta-RAG through evaluation with the SWE-bench Lite dataset. Meta-RAG scores 84.67\% and 53.0\% for file-level and function-level correct localisation rates, respectively, achieving state-of-the-art performance.
\end{abstract}

\end{frontmatter}

%%%%%%%%%%%%%%%%%%%%%%%%%%%%%%%%%%%%%%%%%%%%%%%%%%%%%%%%%%%%%%%%%%%%%%%%

\section{Introduction}
Research into the automation of software development has been at the core of the intersection of AI and software engineering (SE).
Prior work has focused on traditional probabilistic models \cite{hindle_2012} as well as neural-network models \cite{svyatkovskiy2019pythia}.
More recently, code language models \cite{feng_2020codebert,chen_2021evaluating, wang2021codet5} as well as LLMs \cite{li_2023starcoder, rozière_2024codellama} have been utilised in this domain, particularly for the tasks of code generation, code completion, and bug resolution \cite{lozhkov2024starcoder}.

After the introduction of LLMs, due to their superior ability in generating meaningful output with respect to previous AI models, there has been an influx of research on LLM-based software engineering tasks \cite{fan2023large}. As such, LLM multi-agent systems have become one of the standard techniques to implement effective SE automation \cite{park2023generative}. Multi-agent systems enable researchers and industry practitioners to maximise the utility of LLMs by designing specific roles and prompts that achieve modular goals.
Such systems have already been introduced in the software development lifecycle and in AI communities in the form of code generation \cite{yang2024sweagent}, computer control \cite{packer2023memgpt}, and web navigation \cite{zhou2023webarena}, to name a few.

Nevertheless, the software development life cycle is strongly affected by bugs. Therefore, the bug discovery, localisation, and resolution account for a large proportion of software development costs \cite{Jin2023}. Even within this, bug localisation is especially costly and tedious, accounting for approximately 70\% of developers' time spent resolving a bug \cite{muvva2020buglcrosslanguagedataset, Mohsen2022ARO}. This has enticed software engineering researchers and practitioners to develop a number of methods for automating the localisation and repair of software defects, with the latest works utilising LLM agents.

In this paper, we propose Meta-RAG; a Retrieval Augmented Generation (RAG) approach to aid in bug localisation in large codebases. Rather than retrieve the code itself as previous methods have done, we retrieve code meta-data. In order to generate this meta data, we also introduce a novel agent that constructs a compact, natural language representation of a codebase; a codebase summary. Our summaries dramatically reduce the size of codebases by approximately 80\% on average. 

It is vital to measure the performance of our solution using realistic software development scenarios where developers are required to update an existing codebase by adding new features or fixing bugs.
Unfortunately, many of the existing code generation efforts with LLMs utilise benchmarking datasets which are comprised of simple coding examples and are typically contained to one function \cite{jiang2024survey}. Nevertheless, developers are customarily required to update an existing codebase by adding new features or fixing bugs.
Therefore, assessing code implementation systems requires a dataset that requires several sub-tasks (e.g., bug localisation, code generation, code completion, etc.) to be successfully executed.
Consequently, we test our solution against the SWE-bench Lite dataset; a collection of real-world issues on Github from popular open-source repositories. As these are large, complex code repositories,
condensing the codebases into the form of summaries also allows us to overcome two known limitations of LLMs and general AI systems for code implementation: (1) context window length constraints and (2) the diminishing effect of the attention mechanism for long prompts. 

For instance, current GPT models support a 128K context window length (GPT-4o and o1), including input and output tokens \cite{openai_models}. Other LLMs such as Cloud 3-Opus have a 200K token window, and Gemini 1.5 pro has a 128K standard token window. \cite{li2024long}. Nonetheless, a sample code repository in our benchmark, such as SymPY, contains 6,360,381 tokens\footnote{The metric reported here is from a random code commit in this projects history.}; clearly exceeding the input context length of the current common LLMs. For the second challenge of diminishing attention, the LLM requires some form of intervention, such as chain-of-thought or planning, in order to know where and what changes need to be implemented to resolve the issue without risk of hallucination \cite{bairi2024codeplan, agarwal2024codemirage, jiang2023self}, particularly when dependencies are involved. 

Our solution overcomes these challenges through two key contributions; summaries and Meta-RAG. We find that our system achieves the highest successful bug localisation rate at both file and function level amongst state-of-the-art methods presented on the SWE-bench Lite leader board.

\section{Related Work} \label{sec:Related Literature}

Prior to the introduction of LLMs, many researchers have investigated the use of AI for many software engineering tasks, including requirement engineering, software design, code completion, library upgrades, automated testing, bug localisation, automated program repair, and code review \cite{watson2022systematic, alamir2022ai}. After the introduction of LLMs, due to their superior ability in generating meaningful output with respect to previous AI models, there has been an influx of research on LLM-based software engineering tasks \cite{fan2023large}. 
These applications cover a wide range of tasks such as code summarisation and understanding \cite{wang2021codet5, nam2024using, ahmed2022few, sun2024source}, code translation \cite{pan2024lost, eniser2024towards}, code review \cite{alami2025accountability, jensen2024software}, program synthesis \cite{patton2024programming, chen_2021evaluating}, program repair \cite{xia2023automated, jin2023inferfix, bouzenia2024repairagent}, test generation \cite{ryan2024code, liu2024llm}, and planning \cite{tawosi2023search}.

It is clear that most of these applications, however, are focused on the software implementation phase of the software development lifecycle (SDLC), i.e. coding. Nevertheless, research has shown that only 15\% to 35\% of human effort is spent on the implementation phase of the SDLC \cite{meyer_2021}. In this paper we focus on specifically on bug localisation, which remains a challenging task, particularly in large-scale software systems.
Bug localisation is the process of determining the specific location of source code that needs to be modified in order to resolve a specific issue or bug \cite{Takahashi_2021}. Thereby, many previous works have used traditional information retrieval methods in order to aid in tackling this challenge.

BugLocator performs file-level localisation by ranking files based on similarity of the text between the bug report and the source code, taking into consideration information related to historical bugs that have been resolved and using a revised Vector Space Model (rVSM) \cite{zhou2012}. BLUiR can identify bugs at the function-level by utilising source code and bug reports. By structuring and indexing the data, they are able to apply information retrieval methods (combining TF-IDF with BM25) to code constructs, to achieve a higher accuracy for bug localisation \cite{Saha2013}.

Goyal et. al show that the trend of techniques applied for bug triaging has shifted from machine learning based approaches towards information retrieval based approaches \cite{Goyal2017}. Later works began to mix both approaches by incorporating deep learning. One such example is HyLoc, which uses text similarity between bug and source files from rVSM in combination with a Deep Neural Network (DNN) to learn the relationship between the terms in the reports and different code tokens from the source files \cite{lam2015}. DNNLOC \cite{lam2017} works in a similar way with the introduction of the project's bug-fixing history in order to achieve a higher accuracy. 

The emergence of LLMs has increased present research in this space. With this, many works have began implementing Retrieval Augmented Generation approaches (RAG). For instance, RAGFix searches Stack Overflow posts via a Question and Answer Knowledge Graph (KGQA) for bug localisation and program repair \cite{mansur2024}. Nevertheless, one particular benchmarking dataset has seen a rise in popularity to aid in the assessment of the LLM outputs for bug resolution; SWE-bench. SWE-Bench contains 2,294 instances of real-world code issues collected from 12 prominent python repositories on GitHub; the SWE-bench benchmarking dataset \cite{jimenez-2023-swebench}. Each instance of this dataset contains the issue description in natural language, the gold patch which is produced by programmers to resolve the issue, and two sets of unit tests: $pass\_to\_pass$ (unit tests that should pass before and after the change to confirm that the new edit has not broken any previous functionality), and $fail\_to\_pass$ (unit tests that fail due to the reported issue and should pass if the introduced edit resolves the issue successfully). SWE-bench also provide subsets of the original dataset referred to as "Lite" (300 tasks, used in this paper) and "Verified" (500 tasks).

A number of other benchmarking datasets have also been widely used to evaluate software engineering tasks. The HumanEval Dataset (introduced in 2021 by OpenAI), is a curated collection, comprised of 164 handcrafted programming challenges that include language comprehension, basic mathematics, and algorithmic problem-solving \cite{chen_2021evaluating}. MBPP is comprised of 1,000 Python programming challenges that are collected from a diverse set of contributors and are tailored for software engineers at an introductory programming level \cite{austin2021program}. However, both HumanEval and MBPP are comprised of single, isolated functions, and as such, do not provide a realistic simulation of the setting that is required to assess our system for bug resolution.

The SWE-Bench paper not only provides the dataset, but an introduces an approach to bug resolution. This method retrieves code to provide additional context to a prompt through RAG. To achieve this, they utilised a sparse retrieval method, which relies on BM25 and an ``Oracle'' retrieval that returns the files edited by the gold-standard patch referenced in Github \cite{jimenez-2023-swebench}. As an engineer would not have previous knowledge where a new feature should be included, or where a bug is localised, the ``Oracle'' approach is less realistic. 

Another study utilising this dataset provides the LLM with a textual representation of hierarchy of the files in the codebase \cite{xia2024agentless}. With this approach, the LLM is required to decide which files to retrieve based solely on the file name. It narrows down to classes and functions, and line numbers, in iterations, until the LLM localises the lines needed to edit. The performance of this method depends on proper naming of the files in a project, which can vary from project to project. Also, there are cases in which a file hosts multiple classes and functions, and not all functionality included in the file is represented in the file name. 

Agent-based approaches have also been adopted; SWE-Agent \cite{yang2024sweagent} proposed an Agent-Computer Interface (ACI), where pre-programmed tools are provided to an agent. This allows the agent to ideate, search, and retrieve pieces of a codebase in a loop until it finds relevant information to use. Although this approach succeeds in many cases, it may still fail to find relevant information after tens of tries. Our solution builds on the above by utilising an agent-based architecture to create a novel retrieval method that returns relevant code based on summaries.

\section{Methodology} \label{sec:Methodology}

\begin{figure*}[t!]
  \centering
  \includegraphics[width=\textwidth, trim={0 2.5cm 6cm 0.9cm},clip]{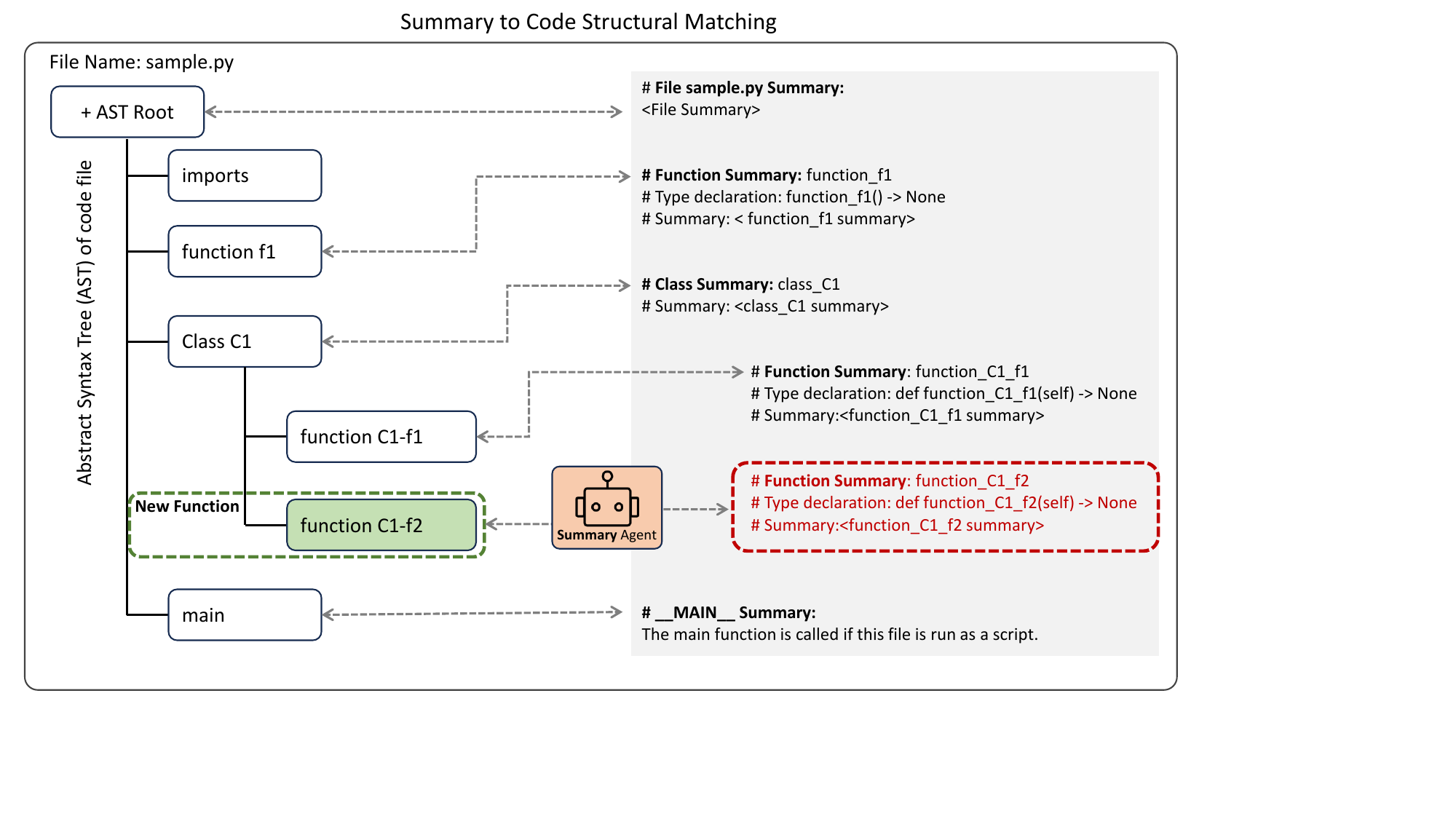}
  \caption{Example of a summary template (right) and a summary update using the AST.}
  \label{fig:summary-update}
\end{figure*} 

\subsection{Summaries}

Bug localisation in a large codebase requires an overall knowledge of the whereabouts of files, classes, and functions, in addition to their connections. To provide our LLM agents with such information, we transform the codebase into a compact yet familiar form of representation for our LLM agents: natural language summaries. 

In this section, we introduce the Summary Agent. The Summary Agent performs two main functions: the generation of summaries and the update of summaries after each task's code is generated by an agent we refer to as the Code Agent. The Code Agent can be any LLM agent (or agent system) that takes a task and outputs a generated code solution. 

Generating summaries for an existing codebase is an offline one-off exercise. We generate and store one summary file per code file. It includes a short summary of the file content (i.e., functionality supported by the code within the file), a short summary of each class or file level function in the file (in the same order and indentation they appear in the file), and the same per any inner class or function in a recursive manner. Each summary item also includes important information about the item: A file summary includes the name and path to the file; a class summary includes the class name and its attributes; and a function summary includes the function name and signature. There may still exist some code that is not a class or function; this is then summarised into "\_\_MAIN\_\_". Note that imports are not included in the summary as they are present in the code that will later be augmented with the LLM prompt. An example of a summary template is shown in the grey box of Figure \ref{fig:summary-update}.

The code file's content is used in combination with proper instructions in a prompt to ask the LLM to generate a summary. 
The generated summaries are stored in a database along with their code files and are structurally matched to their corresponding file structures by parsing into an abstract syntax tree (AST), which is a common data structure used to represent the structure of a program. 
This is carried out by parsing the generated summary and identifying the code elements, with their order and hierarchy in the file. We traverse the AST node-by-node, marking the matching summary section to each of the main AST nodes (i.e. functions and classes). We match then those with the AST of the code file, to make sure: a) all code elements (classes and functions) have corresponding summaries, and b) they are in the correct order and hierarchy. If any misalignment is identified, the issue is fixed by reordering the summary elements or updating the summary if needed. 

After this matching procedure, the summaries and code content are stored in a data structure similar to that of an AST, with each code element stored alongside its summary. This helps with easier code and summary traversal and retrieval, especially when we want to return the summary for a given code element or a code element for a given summary. Figure \ref{fig:summary-update} better depicts this mapping. By compressing the codebase into this new representation, we observe a 79.8\% ($std.$ 9.1\%) reduction on average for the SWE-bench Lite code repositories. 

Summary agent is also responsible for keeping the code summaries up to date with changes to the codebase.
Once a new function is introduced or edited by the Code Agent, we generate a new summary for the new element using the Summary Agent, and update the summary file by injecting the summary into the corresponding location, taking the structural matching into consideration. This is depicted again in Figure \ref{fig:summary-update} where \textit{function C1-f2} is the new function created for a task, and the summary agent is able to inject the summary for this function into the correct location in the database.

\begin{figure*}[t!]
  \centering
  \includegraphics[width=\textwidth, trim={0 5.5cm 4.5cm 0},clip]{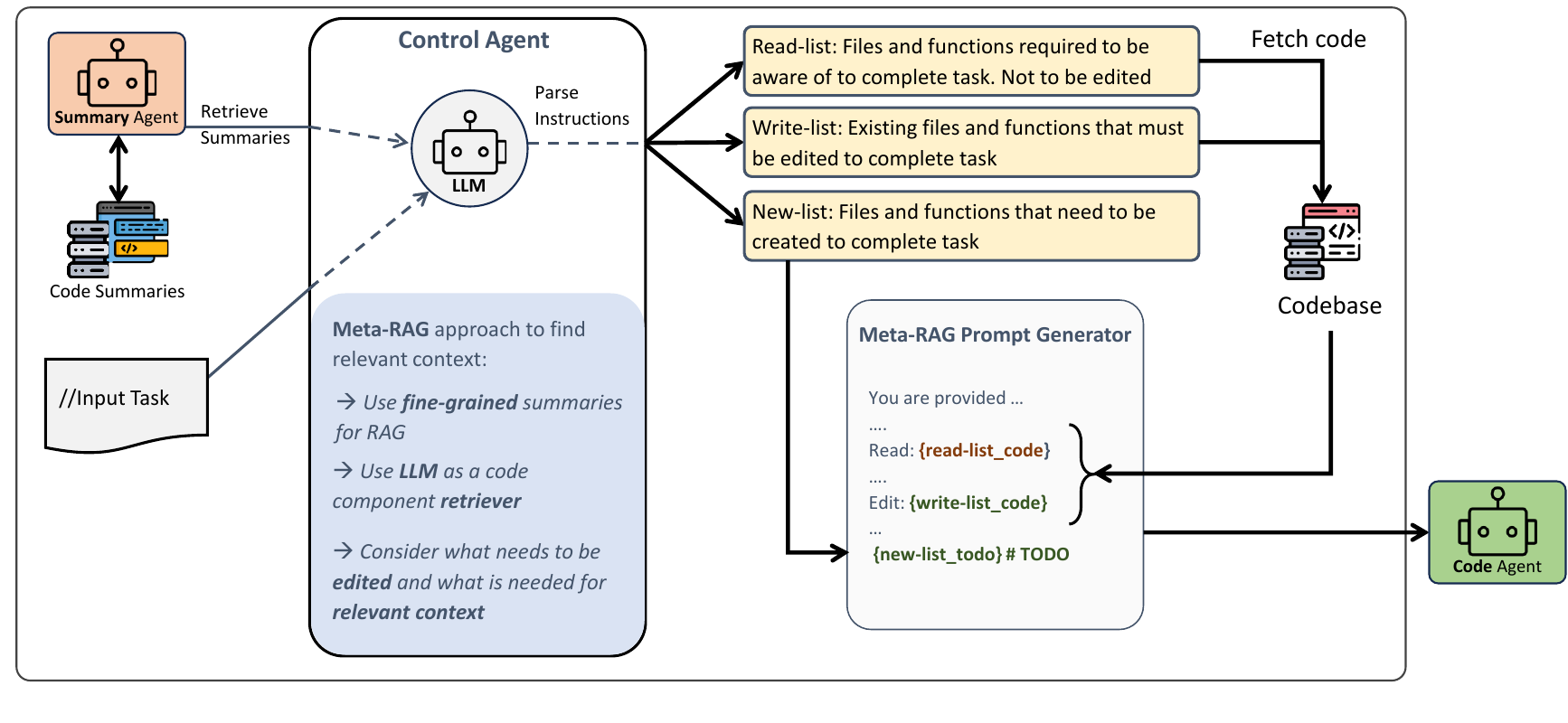}
  \caption{Meta-RAG architecture diagram.}
  \label{fig:srag-arch}
\end{figure*}

\subsection{Meta-RAG}

Once the summaries are generated and stored, LLM agents will work with the summaries instead of code to localise the change. This will help overcome the second challenge for working with large codebases: the diminishing attention effect.
On that account, we introduce Meta-RAG; a novel retrieval method that uses the summaries of the code units from Summary Agent at different hierarchical levels, and prompts the LLM to identify the relevant pieces of code required from the codebase to complete the task at hand. The advantage of this method, in combination with the summaries, is that there are different levels of granularity that the LLM can decide to retrieve (file, class, function), typically starting at a higher level (file) and honing in on the location at a lower level (function). The LLMs are able to more efficiently traverse through the codebase, as the summaries convey more information in fewer tokens. When code augmentation or bug resolution is intended, we first need to find the location within the codebase that needs to be changed or extended to support a bug fix or a new feature. In the case of the SWE-bench dataset, the tasks are predominantly bugs; thus this component aids particularly in bug localisation. 

Working with a large codebase means there are many interdependencies within the code in the form of API calls. Thus the new code should attempt to reuse available code via these API calls, rather than re-implementation. This is possible with Meta-RAG.
Via prompting, an agent referred to as Control Agent, asks the LLM to find the code units within the summaries that are relevant to the task at hand, and return them in three lists: \textit{Read-list}, \textit{Write-list}, and \textit{New-list}. The agent instructs the LLM to include into the \textit{Read-list}, the code units that provide context or are useful for implementing the change. The \textit{Write-list} contains the code units that need to be amended to reflect the new change. And the \textit{New-list} may include new functions, classes, or files that the LLM needs to author for a new feature. 
In the case of a new project where there are no summaries, all code would be in the \textit{New-list}. 
This approach encourages the LLM to reuse previous code in the codebase and helps with a natural integration of the new code into the existing codebase \cite{wang2024and}. Furthermore, previous RAG approaches rely solely on identifying the optimal \textit{Write-list}. We propose that a \textit{Read-list} is provided to supply additional context to the LLM and aid in downstream resolution of the task instances. 

The aforementioned retrieval instructions are then sent to a Code Agent. The Code Agent should retrieve the specified pieces of code from the existing codebase and augment these into a prompt, in combination with the task, in order to generate a code patch. The LLM output is parsed to extract the code snippets with regular expression and rule based parsing to determine the filename and function name to assess localisation. An overview of the system architecture is exhibited in Figure \ref{fig:srag-arch}.

There still exist repositories large enough that their summaries do not fit within the context window of the LLM. Consequently, Meta-RAG starts with file-level summaries (which are one-liner file summaries) and the Control Agent requires the LLM to short-list the files that are relevant to the current task. Once the files are selected, the Control Agent is provided with the full summary of the selected files and asked to select the relevant code parts (classes and functions). This process can be broken down into many more rounds of retrieval if any of the context gets larger than the context-window, or a pre-set limit. Again, this remains possible due to availability of summaries in different hierarchical levels.

\begin{table*}[h]
\caption{The Total Code Tokens and Summary Tokens required to be generated for instances in each repository within the SWE-bench Lite dataset is presented with the Codebase Size Reduction rate achieved via summarisation. Columns on the right half of the table show the reduction rate we achieved with summary update method: New Commit Tokens shows the total tokens of code change in the consecutive instances of a repo, Updated Summary is the number of the summary tokens generated for the code changes, and Saved by Summary Update shows the rate of codebase size reduction using summary update. Also Saved on Code Submission shows the reduction rate in changed code tokens submitted for summarization compared to re-summarising the codebase for each task instance.}
\centering
\label{tbl:summary_tokens_saved}
\resizebox{\textwidth}{!}{%
\begin{tabular}{l r r r r | r r r r}
\toprule
& &Total Code& Total Summary & Codebase Size& New Commit & Updated Summary & Saved on& Saved by\\
Repo & \# Instances &Tokens & Tokens &Reduction \%& Tokens & Tokens & Code Submission \% & Summary Update \% \\
 \midrule
 astropy & 6 & 9,259,099 & 1,632,048 & 82.4\% &4,020,211 & 675,685  &58.6\% & 83.2\%\\
 django & 114 & 119,282,807 & 46,566,608 & 61.0\% & 25,200,201& 8,624,366  & 81.5\%  & 65.8\% \\
 flask & 3 & 208,232 & 48,165 & 76.9\% & 113,880& 24,428 & 49.3\% & 78.5\%\\
 matplotlib & 23 & 30,621,968 & 4,059,709 & 86.7\% & 9,722,932 & 1,058,121  &73.9\% & 89.1\%\\
 pylint & 6 & 2,091,552 & 529,337 & 74.7\%  & 1,025,400& 232,386  & 56.1\% & 77.3\%\\
 pytest & 17 & 3,332,021 & 1,086,032 & 67.4\%  & 2,192,246 & 689,091& 36.5\% & 68.6\%\\
 requests & 6 & 2,238,204 & 193,130 & 91.4\% & 375,410& 98,343  & 49.1\% & 73.8\%\\
 scikit-learn & 23 & 26,031,761 & 3,580,341 & 86.3\% &8,297,394 & 1,116,840 & 68.8\% & 86.5\% \\
 seaborn & 4 & 1,022,276 & 137,697 & 86.5\%  & 397,902& 63,110 & 54.2\% & 84.1\%\\
 sphinx & 16 & 10,037,071 & 2,398,412 & 76.1\%  & 4,437,583 & 1,130,412 & 52.9\% & 74.5\%\\
 sympy & 77 & 399,640,756 & 45,324,399 & 88.7\% & 81,509,674& 10,373,951  & 77.1\% & 87.3\%\\
 xarray & 5 &  1,859,380 & 389,706 & 79.0\% & 1,327,671& 245,474  & 37.0\% & 81.5\%\\
  \midrule
 Total& 300 & 605,625,127& 105,945,584 & 79.8\% (Mean) & 138,620,504 & 243,32,207 & 57.9\% (Mean)  & 79.2\% (Mean)\\
\bottomrule
\end{tabular}}
\end{table*}

\section{Experimental Setup}

We evaluate our approach using the SWE-bench Lite dataset \cite{swelite}. This dataset is a sub-sample of the full SWE-bench dataset containing 300 instances across the 12 repositories. SWE-bench Lite is sampled by the authors of the original dataset to be more self-contained, with a focus on evaluating functional bug fixes.

To reduce the cost of summarisation, we only summarise the earliest instance for each repository (based on their creation date) and update the summaries for instances present after that. We do this by getting a git diff between the consecutive instances and updating the summaries for the changed parts in the code. Creating a summary of the entire codebase for each task instance, rather than updating summaries would require over 605 million code tokens to be provided to the LLM (See Table \ref{tbl:summary_tokens_saved}). 

With regards to the LLM, for all of the experiments, we use GPT-4o. This LLM model has a 128K context length. When comparing our approach with other solutions, we control for the LLM by comparing our model to six top approaches on the SWE-bench Lite leaderboard, filtered only to open source projects using GPT-4o \footnote{These metrics reflect the leaderboard as of 5 May, 2025.}.

We assess and report the localisation rate, which shows the proportion of instances where our tool was able to identify the location of the fix for the problematic units of code successfully. We compute this metric by comparing the agent-identified locations to fix, with those edited by the ``gold'' patch at both the file level and the function level (See Equation \ref{eq:rate}).
We rely on the lines changed by the developer who has fixed the bug as the ``gold'' standard. Parsing the available git patch in the benchmark, we extract the edited lines and use the code AST to find the corresponding edited function from the source file. 
To compare our localisation metric to that of previous works, we calculate the same metric on their outputs reported in the SWE-bench git repository \cite{sweevaluations}. 

 \begin{equation}
     g(t) = 
     \begin{cases}
      1 & \text{if change(Meta-RAG Write\_List}_t\text{)= change(gold-patch}_t)\\
      0 & \text{otherwise}
    \end{cases}   
\end{equation}

\begin{equation}
\label{eq:rate}
    \% Correct Localisation = \frac{100}{|dataset|} \sum_{task \in dataset}{g(task)}
\end{equation}

As a baseline, we also compare our approach to BM25-based retrieval methods. Our baselines include BM25 from Lucene index search\footnote{We used {\tt pyserini} python implementation for Lucene index search with BM25.} and BM25-Plus \cite{lv2011lower}\footnote{We used {\tt rank-bm25} python implementation for BM25-Plus.}. BM25 is a popular ranking function used in information retrieval to estimate the relevance of documents to a search query. It works by calculating a score for each document based on the appearance of the query terms in it. It uses both term frequency and inverse document frequency for this calculation. Equation \ref{eq:bm25} shows how the BM25 score is calculated:

\begin{equation}
\label{eq:bm25}
    s_q = \sum_{t \in q}{log \left(\frac{N- df_t +0.5}{(df_t + 0.5)}\right)\frac{(k_1+1).tf_{td}}{k_1.\left((1-b) + b.(\frac{L_d}{L_{avg}})\right)+tf_{td}}}
\end{equation}

\noindent where the retrieval score ($s$) for a given multi-term query $q$, is calculated as the sum of individual term ($t$) scores. In this equation, $N$ is the number of documents in the collections, $df_t$ is the number of documents containing the term (the document frequency), and $tf_{td}$ is the number of times term $t$ occurs in document $d$. $L_d$ is the length of the document (in terms) and $L_{avg}$ is the mean of the document lengths. The $log$ term calculates the inverse document frequency for term $t$. There are two tuning parameters, $b$, and $k_1$, set to their default values 0.75 and 1.5, respectively.

BM25-Plus improves on the original BM25 algorithm by lower-bounding the contribution of a single term occurrence by adding a $\delta$ term to the fraction in Equation \ref{eq:bm25}, to avoid penalisation by long documents. This helps a single occurrence of a search term to contribute at least a constant amount to the retrieval score value, regardless of document length. The $\delta$ is set to 1.0 by default.

The BM25-Plus also uses Robertson-Walker IDF \cite{lee2007idf} (i.e., $log(\frac{N+1}{df_t})$), which tends to zero as $df_t$ tends to $N$, instead of a common Robertson-Sparck Jones IDF \cite{jones2000probabilistic} used in Equation \ref{eq:bm25}, which for a 1-term query containing a term that occurs in more than half the documents, ranks all documents not containing that term higher than those that do. This change makes BM25-Plus always consider documents containing the term to be more relevant than those that do not.

We use BM25-Plus retrieval in two modes. First, we use it to retrieve full code files (i.e., each code file is treated as a document) based on their relevance to the bug report, until we fill a pre-set context limit. Specifically, we use BM25-Plus to rank all code files in the codebase by their relevance to the bug report, and include the most relevant files in the context until we reach to the pre-set context limit  (i.e., 13K, 27K, 50K, and 80K tokens). Then, providing only these files, we use an LLM (GPT-4o) to find the file and function where the bug fix should apply. This is a similar approach used by SWE-bench RAG mechanism~\cite{jimenez-2023-swebench}. Furthermore, we also use BM25 directly on code to retrieve bug locations. To do this, using code AST, we divide code files into separate functions and index them using BM25 as separate documents. For each code file, we also include any code outside a function or class in a single document called ``MAIN''. Then we use BM25 to retrieve the function (or MAIN part) which is relevant to the bug report. We run this baseline using both BM25 using Lucene search and BM25-Plus.

\begin{table*}[t]
\caption{Correct Localisation rate (\%)  on SWE-bench Lite benchmark. Listed are Information Retrieval (IR) and Software Engineering (SE) approaches.}
\centering
\label{tbl:results}
\resizebox{\textwidth}{!}{%
\begin{tabular}{l l r r | l l r r} 
\toprule
\multirow{2}{*}{Approach (IR)} & \multirow{2}{*}{LLM} & \multicolumn{2}{c}{\% Correct Localisation}&\multirow{2}{*}{Approach (SE)} & \multirow{2}{*}{LLM} & \multicolumn{2}{c}{\% Correct Localisation}\\
& & File & Function & & & File & Function\\
\midrule
BM25-Lucene & - &33.67\%&13.00\%&Aider \cite{aider} & GPT-4o & 65.33\%& 32.67\%\\
BM25-Plus & - & 19.19\%& 6.40\%&Agentless 1.5 \cite{agentless15}& GPT-4o & 69.67\% & 37.00\%\\
BM25-Plus (13K) &GPT-4o&38.00\%& 13.33\%&Agentless RepoGraph \cite{repograph} & GPT-4o& 71.00\% & 36.00\%\\
BM25-Plus (27K) &GPT-4o&54.18\%& 24.08\%&AppMap Navie v2 \cite{appmappnavi}&GPT-4o&56.67\%&26.00\%\\
BM25-Plus (50K) &GPT-4o&60.00\%& 25.33\%&AutoCodeRover \cite{zhang2024autocoderover} & GPT-4o & 65.00\% & 33.00\%\\
BM25-Plus (80K) &GPT-4o&57.73\%& 25.77\%&ReproducedRG \cite{reproducedRG}&GPT-4o&71.67\%&39.33\%\\
\midrule
&&&&Meta-RAG&GPT-4o&\textbf{84.67}\%&\textbf{53.00}\%\\
\bottomrule
\end{tabular}}
\end{table*}

\section{Results}
The results of percentage correct localisation at file and function level are presented in Table \ref{tbl:results}. We provide a comparison with the open-source models that achieve state-of-the-art performance on the SWE-bench Lite dataset, using GPT-4o as the LLM. The top six models from the leaderboard are presented, and we can see that our correct localisation rate is the highest among all methods at both file-level (with 84.67\%) and function level (with 53.0\%). 

The results of the baseline study show that BM25 is able to identify the correct files with a rate of 60.0\%, in the best case, with 50K context window, and the correct functions with a rate of 25.77\% with 80k context window. Using BM25 independent of an LLM leads to a lower score (33.67\% and 19.19\% for Lucene-based and BM25-Plus, respectively, for file-level and 13\% and 6.4\% for function-level).

Furthermore, we have observed an average code reduction of 79.8\%, ranging between 91\% (requests repository) and 61\% (django repository) among the 12 SWE-bench repositories. Table \ref{tbl:summary_tokens_saved} shows the total number of code token alongside the number of summary tokens generated for a sample instance from each of the 12 repositories in the SWE-bench dataset.

\begin{table}
\caption{The average Time taken, tokens used, and cost --based on August 2024 OpenAI pricing list-- to resolve instance of each repository (Repo) from SWE-bench Lite.}
\centering
\label{tbl:cost}
\resizebox{\columnwidth}{!}{%
\begin{tabular}{lrrr}
\toprule
	&	Time Taken	&	Tokens	&	Cost\\
Repo	& (seconds)	& Used	& (USD)\\

\midrule
astropy	&	51.69	&	21,429.63	&	1.29	\$ \\
django	&	62.32	&	33,408.65	&	2.00	\$ \\
flask	&	25.60	&	5,320.67	&	0.32	\$ \\
matplotlib	&	52.15	&	16,140.09	&	0.97	\$ \\
pylint	&	36.58	&	13,292.83	&	0.80	\$ \\
pytest	&	62.94	&	15,033.59	&	0.90	\$ \\
requests	&	45.97	&	8,718.83	&	0.52	\$ \\
scikit-learn	&	48.36	&	14,626.73	&	0.88	\$ \\
seaborn	&	38.80	&	8,177.00	&	0.49	\$ \\
sphinx	&	61.68	&	12,398.44	&	0.74	\$ \\
sympy	&	137.29	&	32,026.42	&	1.92	\$ \\
xarray	&	78.95	&	13,284.80	&	0.80	\$ \\
\midrule
Mean & 58.53 & 16,154.81 & 0.97 \$\\
\bottomrule
\end{tabular}}
\end{table}

\section{Discussion}

Our results in Table \ref{tbl:summary_tokens_saved} show that we are able to compress the repositories to achieve a reduction of 79.8\% on average. The maximum compression was the requests library at 91.4\% (a smaller code repository) and the minimum was django at 61\% (one of the larger repositories). Initial analysis showed that this is a result of repositories with more specific domain knowledge requiring more tokens even after summarisation. 

Furthermore, by implementing the summary update approach onto the SWE-bench dataset, we observe a 57.9\% reduction in tokens compared to re-summarising the codebase for each task instance, making this approach both a computationally efficient and cost-effective approach. The real-time compression per updated instance is 79.2\%; comparable to the one-off compression of the entire codebase. 

The process of generating summaries for an existing codebase may initially introduce a one-time overhead cost. However, we show that the strategic use of these summaries can lead to substantial reductions in the ongoing costs of utilising LLMs over time. By leveraging summaries, developers can streamline the integration of LLMs into their workflow, ultimately enhancing efficiency and minimising resource expenditure in the long run. This shows how summary updates could be efficient for a code base in the long run.

Employing codebase summaries within a prompt, rather than submitting the actual code, can offer an additional valuable advantage in safeguarding the security of proprietary code. This approach circumvents the need to submit source code to a large language model (LLM) for various software development tasks, thereby protecting sensitive information. This security measure is particularly effective when the summaries are generated and regularly updated by an internal LLM, ensuring that the proprietary code remains confidential and secure. 

Additionally, these generated summaries serve a dual purpose: they can enhance existing code documentation, providing additional context and clarity, and they can assist developers in gaining a deeper understanding of the codebase. By offering concise and informative insights, summaries facilitate a more efficient and secure development process, empowering developers to work with greater confidence and comprehension alongside, especially alongside AI agents.

The results then display the results of the assessment of localisation at the file level and the function level. We show that the Meta-RAG approach achieved the highest correct file-level (84.67\%) and function-level (53.0\%) localisation rate amongst the top performing GPT-4o approaches. The BM25 retrieval baselines were able to localise the bugs to a reasonable file level (60\%) and function level (25.77\%) accuracy, closing in on the more advanced approaches. We observe that in the case of file level, as we increase the context length, the accuracy increases. However, once the context length increased from 50k to 80k, the localisation correctness drops. This confirms the diminishing attention effect that our approach aims to overcome.

Our approach also reduces the cost of using LLMs for software development, particularly within a real-world continuous development environment. The price in dollars for running one task through our system ranges from 0.32\$ to 2.00\$, with the mean price at 0.97\$. This depends not only on repository size, but file sizes as they would require more iterations of traversal through summary hierarchies for localisation. We also record the time taken to pass one task through the system with the Meta-RAG framework. This ranges from 25.60 seconds to 137.29 seconds, with the mean being 58.53 seconds. As expected, the longest time corresponds to sympy, the largest repository, containing approximately 6 million tokens.

Overall, the file-level and function-level correctness in addition to the cost benefits is a remarkable outcome as bug localisation is a time-consuming task \cite{polisetty2019usefulness}. Identifying the correct file can save developers' time in bug fixing, particularly in cases where the codebase is large or the developer is inexperienced with it. Thus achieving a high performance on this metric goes a long way towards a practical solution in industry. This promising result for localisation can aid in improving the bug resolution rates for LLM agents, something we plan to investigate in future work. Finally, achieving a high localisation capability for bug resolution is a proxy to understanding the usefulness of code summaries and our Meta-RAG approach in helping LLMs code independently in the future.

\section{Conclusion and Future Work}

In this paper, we introduce a multi-agent LLM-based framework designed to utilise information retrieval methods for bug localisation. We evaluated the system's capabilities on the popular SWE-bench Lite benchmark, achieving state-of-the art file-level and function-level correct bug localisation rates. This is as a result of the use of code summaries for context retrieval and Meta-RAG, both introduced in this paper. We also showed that summarisation of the codebase reduces its size in tokens by approximately 80\% in average. This reduces the cost of using LLM-based agents for software development down-stream tasks that require code retrieval in the long run.

The configuration of the agents was set up to tackle bug localisation, in addition to limited context window length, and diminishing attention challenges. In the future, we plan to evaluate and build on this work for the downstream task of bug resolution and even new feature implementation.

\section*{Disclaimer} 
This paper was prepared for informational purposes by the Artificial Intelligence Research group of JPMorgan Chase \& Co. and its affiliates ("JP Morgan'') and is not a product of the Research Department of JP Morgan. JP Morgan makes no representation and warranty whatsoever and disclaims all liability, for the completeness, accuracy or reliability of the information contained herein. This document is not intended as investment research or investment advice, or a recommendation, offer or solicitation for the purchase or sale of any security, financial instrument, financial product or service, or to be used in any way for evaluating the merits of participating in any transaction, and shall not constitute a solicitation under any jurisdiction or to any person, if such solicitation under such jurisdiction or to such person would be unlawful.

%%%%%%%%%%%%%%%%%%%%%%%%%%%%%%%%%%%%%%%%%%%%%%%%%%%%%%%%%%%%%%%%%%%%%%%%

%%% Use this command to include your bibliography file.

\bibliography{references}

\end{document}